\documentstyle[12pt]{article}

\def\be{\begin{equation}}
\def\ee{\end{equation}}
\def\ba{\begin{array}{c}}
\def\ea{\end{array}}

\def\ben{$$}
\def\een{$$}
\begin{document}

\titlepage
%\vspace*{2cm}

\begin{center}{\Large \bf
Non-Hermitian matrix description of the ${\cal PT}-$symmetric
anharmonic oscillators }\end{center}

\vspace{5mm}

\begin{center}
Miloslav Znojil
\vspace{3mm}

\'{U}stav jadern\'e fyziky AV \v{C}R, 250 68 \v{R}e\v{z},
Czech Republic\\

\end{center}

\vspace{5mm}

\section*{Abstract}

The ${\cal PT} -$symmetric differential Schr\"{o}dinger equation
$H\psi=E\psi$ with the operator $H=H(x) = p^2 + a\,x^4 +
i\,\beta\, x^3+c\,x^2+i\,\delta\,x \equiv H^*(-x)$ on
$L_2(-\infty,\infty)$ is studied. At $a > 0$ it is re-arranged as
a linear algebraic diagonalization. With rigorous proof, our
non-variational construction of bound states offers an
infinite-dimensional analogue to the recent finite-dimensional
quasi-exact solution available at the less common $a<0$.

\vspace{9mm}

\noindent
 PACS 03.65.Ge,
03.65.Fd

%\vspace{9mm}

%\begin{center}
%{\small \today, ahopt.tex }
%\end{center}

\newpage

\section{Introduction}

The study of the general parity-breaking anharmonic oscillators
 \be
V(x)=a\,x^4+b\,x^3+c\,x^2+d\,x \label{AHO}
 \ee
has a colourful and inspiring history \cite{Weniger}. Its latest
turn came with the recent letter by Bender and Boettcher
\cite{BBjpa} who discovered that after a partial restoration of
symmetry in the complex plane,
 \be
V(x)=V^*(-x),
  \label{PT}
  \ee
these potentials may become solvable quasi-exactly
\cite{Ushveridze}.  The extensive numerical experiments indicate
that all the spectrum of energies $E$ in the similar potentials is
real, bounded and discrete. According to the conjecture by Daniel
Bessis \cite{Bessis} this puzzling observation might be a
straightforward mathematical consequence of the ``weakened
hermiticity" (\ref{PT}). With motivations ranging from field
theory \cite{Meis} and nuclear structure \cite{Navr} up to solid
state physics \cite{Hatano} this hypothesis finds its further
support in a few explicit analytic \cite{Junker} and numerical
\cite{Meisinger} constructions and semi-classical \cite{BB} or
perturbative \cite{mytri} arguments as well as in several
available rigorous mathematical proofs \cite{Caliceti}.

In the particular model (\ref{AHO}) the condition (\ref{PT}) means
that the couplings $a$ and $c$ remain real while their partners
$b=i\,\beta$ and $d=i\,\delta$ are purely imaginary.  The
attention of paper \cite{BBjpa} was solely paid to the {\em
negative} values of the asymptotically dominant coupling $a$
because of the related finite-dimensional reducibility and
subsequent partial solvability of the bound state problem at
certain special couplings and energies. In the present paper we
intend to complement and complete the latter study by a parallel
analysis of its ``less solvable" alternative with $a>0$. One has
to imagine that in spite of the manifest non-hermiticity of the
related Hamiltonian the procedure of quantization may be kept
equally well defined at any sign of $a$ \cite{Turbiner}. Extensive
discussions of this point date back to the famous Dyson's argument
(\cite{Dyson}; cf. the very recent summary in \cite{BBqed}). The
same or similar formalism covers even the limiting case with the
vanishing $a=0$ at any complex $b\neq 0$ \cite{Alvarez}.

The main reason for the conventional $a>0$ is the simplicity of
its physical interpretation. In an unquantized world the minus
sign of $a$ would mean that the particle can disappear and return
from infinity in a finite time. In comparison, the asymptotically
real and growing $V(x)$ admits a more immediate intuitive
understanding. Its choice weakens the impact of the unusual
invariance (\ref{PT}) which, in certain applications, mimicks the
combined effect of parity and time reversal ${\cal PT}$ \cite{BB}.
With $a>0$ we also get in a closer contact with the already
existing calculations \cite{list} and with the re-summations of
perturbation series, say, in terms of the so called Hill
determinants \cite{Ginsburg,Estrin} or analytic \cite{Singh} and
matrix \cite{Graffi} continued fractions. The new, non-Hermitian
options $\beta\neq 0$ and $\delta\neq 0$ open the new
perspectives.

We intend to show that the positivity of $a$ which excludes the
quasi-exact solvability need not contradict an efficient linear
algebraic description of bound states. We shall re-write the
differential Schr\"{o}dinger equation in an equivalent matrix
form. Although its dimension remains infinite, its structure and
derivation will parallel its Hermitian Hill-determinant
predecessors characterized globally by a loss of their hermiticity
({\it pars pro toto}, the reader may consult the review
\cite{Classif}). For our present non-Hermitian interaction
(\ref{AHO}) + (\ref{PT}) such a loss is much less harmful.

\section{Non-terminating recurrences at $a>0$}

Forces (\ref{AHO}) with the ``weak" symmetry (\ref{PT})
enter the differential Schr\"{o}dinger equation
 \be \left
(-\,\frac{d^2}{dx^2} + a\,x^4+i\,\beta\,x^3+c\,x^2+i\,\delta\,x
 \right )
 \, \psi(x) =
E  \, \psi(x), \ \ \ \ \ x \in (-\infty,\infty)
 \label{SE}
 \ee
which has the two independent asymptotic solutions
 \be
\psi^{(\pm)}(x) = \exp \left [ u\frac{x^3}{3} + v\frac{x^2}{2} +
{\cal O}(x)
  \right ], \ \ \ \ \ \ \ u
=\pm\sqrt{a}\neq 0, \ \ \ \ \
v=i\frac{\beta}{2u}.
  \label{asymp}
 \ee
This explains the difference between $a<0$ and $a>0$. In the
former case we may move the real axis downwards in the complex
plane, $ x = r-i\eta, \ \eta > 0,\ r\in (-\infty ,\infty )$.
Whenever we pick up $ \eta>-\beta/4\sqrt{|a|}$, we discover that
the asymptotic solution $\psi^{(-)}(x)$ remains integrable at {\em
both} the ends of the real axis $r$. Some of the (necessarily,
analytic) bound states $\psi^{(exact)}(x)$ may (and do) acquire an
elementary form of an exponential-times-polynomial product for
$a<0$ \cite{BBjpa}.

Let us now consider a positive coupling $a>0$ and re-scale its
value to $a=1$ for simplicity.  The general solution of our
Schr\"{o}dinger equation (\ref{SE}) will have the form
$\psi^{(gen)}(x) = \mu_{\pm}\,\psi^{(+)}(x)
+\nu_{\pm}\,\psi^{(-)}(x)$. It may only be made compatible with
the required asymptotic decrease near $x \to \pm \infty$ by the
sign-of-$x$-dependent choice of its parameters,
 \be \ba
\psi^{(phys)}(x) = \nu_{+}\,\psi^{(-)}(x), \ \ x \gg 1,\\
\psi^{(phys)}(x) = \mu_{-}\,\psi^{(+)}(x),
 \ \ x \ll - 1.
 \ea
\label{genere}
 \ee
In contrast to the preceding case, the exact solution cannot be
constructed as a product of an exponential with a polynomial
anymore. One has to resort to the next eligible possibility, say,
 \be
\psi^{(ansatz)}(x)=e^{-sx^2}\,\sum_{n=0}^{\infty}\,h_n\,(ix)^n.
 \label{ansatz}
 \ee
This is a manifestly ${\cal PT}-$invariant infinite-series ansatz.
Abbreviating $ix=y$ we derive the recurrences
 \be
A_n\,h_{n+2} +C_n\,h_n + \delta \,h_{n-1} +
\theta\,h_{n-2}-\beta\,h_{n-3} + h_{n-4}=0. \label{recurrences}
 \ee
All its coefficients are real, $A_n=(n+1)(n+2)$, $C_n=4sn+2s-E$
and $\theta = 4s^2-c$. By construction, the set
(\ref{recurrences}) is equivalent to our differential equation
(\ref{SE}). At all the tentative energies $E$ it defines the
coefficients $h_n$ from an input pair $h_0$ and $h_1$.  We have
to determine these parameters via a fit of (\ref{ansatz}) to the
appropriate boundary conditions
\be
 \psi^{(ansatz)}(X_R) = 0= \psi^{(ansatz)}(-X_L), \ \ \ \ \ \
 X_R \gg 1, \ \  X_L \gg 1.
 \label{bc}
 \ee
In comparison with the other numerical methods of solution of the
bound state problems with symmetry (\ref{PT}) \cite{Meisinger}
such a recurrently specified recipe does not look any superior,
especially because it requires a cumbersome numerical limiting
transition $X_{R,L} \to \infty$. A deeper insight and
simplifications are asked for.

\section{The asymptotics of coefficients $h_n$ at $n \gg 1$}

Recurrences (\ref{recurrences}) form a linear difference equation
of the sixth order. One may recall the standard theory of its
solution \cite{Birkhof} as well as its immediate application to
quartic oscillators \cite{Karel,Karl}. At the large indices $n$,
the sextuplet of the independent asymptotic solutions $h_n$
acquires the general Birkhoff form as presented, say, in ref.
\cite{soukr}. All these solutions decrease as $h_n \sim {\cal
O}(n^{-n/3})$ at least. For our present purposes, they may easily
be re-derived as follows.

Firstly, in the leading-order approximation, we replace equation
(\ref{recurrences}) by the mere two-term dominant relation between
$h_{n+2}$ and $h_{n-4}$. This inspires us to change variables $h_n
\to g_n$ and we re-write all the six independent solutions in the
same compact form
\be
h_n(p)=\frac{\lambda^n(p)\,g_n(p)}{(3^{1/3})^n\,\Gamma(1+n/3)}, \
\ \ \ \ \ \ \ \ \ \   \ \ \
 p = 1, 2, \ldots, 6.
 \label{coeff}
 \ee
The $p-$dependent complex parameter $\lambda(p) =\exp
[i(2p-1)\pi/6]$ characterizes the dominant $n-$dependence of the
separate solutions while the new functions or coefficients
$g_n=g_n(p)$ vary more slowly with $n$.

This confirms the linear independence of our six solutions but
leaves their absolute values indistinguishable. In order to remove
this degeneracy in size we re-introduce equation
(\ref{recurrences}) in its amended, second-order asymptotic form
\be
 g_{n+2} - g_{n-4} = \frac{4s\lambda^4}{n^{1/3}}\,g_n
 - \frac{\beta\,\lambda}{n^{1/3}}\,g_{n-3} + {\cal O}
 \left ( \frac{g_n}{n^{2/3}} \right ) .
 \label{cond}
 \ee
With most of the components of the Stirling formula still hidden
within the error term, the smallness of the ratio $1/n^{1/3}$
enables us to infer that
 \be
 g_n= e^{\gamma \,n^{2/3}+{\cal O}(n^{1/3})}, \ \ \ \ \ n \gg 1.
\label{second}
 \ee
The complex exponent $\gamma =\gamma(p) = s \lambda^4(p)-
\beta\,\lambda(p)/4$ depends on $p$. An elementary trigonometry
gives the explicit formulae
 \ben
{\rm Re}\ \gamma (1)={\rm Re}\ \gamma (6) = -\frac{\sqrt{3}}{8}
\beta - \frac{s}{2}, \ \ \ \ \
 {\rm Re}\ \gamma (2)={\rm Re}\ \gamma
(5) = s,
 \een
 \be
  {\rm Re}\ \gamma (3)={\rm Re}\ \gamma (4) =
\frac{\sqrt{3}}{8} \beta -
 \frac{s}{2}.
 \label{estim}
  \ee
In combination with eq. (\ref{coeff}) this already implies that
the radius of convergence of our Taylor series (\ref{ansatz}) is
infinite. The function $\psi^{(ansatz)}(x)$ is unique and well
defined at any complex $x$. Its shape is fully determined by the
energy $E$ and by a not yet specified choice of the two initial
complex coefficients $h_0$ and $h_1$.

The second important consequence of identities (\ref{estim}) is
that whenever we satisfy the condition
\be
 s > \frac{| \beta |}{4\sqrt{3}}
\label{constr}
 \ee
the general solution $ h_n=\sum _{p=1}^6 G_p\,h_n(p)$ itself will
be asymptotically dominated by its two most quickly growing
components,
\be
 h_n=G_2\,h_n(2)+G_5\,h_n(5), \ \ \ \ \ \ \ n \gg 1.
\label{gensol}
  \ee
In this sense we are free to set $ G_1=G_3=G_4=G_6=0$ in the
asymptotic domain of $n \gg 1$. Each choice of the energy $E$ and
initial $h_0$ and $h_1$ will only generate a different, $x-$ and
$n-$independent pair of coefficients $G_{2}$ and $G_{5}$.

\section{The asymptotics of $\psi^{(ansatz)}(x)$ at $|x| \gg 1$}

Equation (\ref{gensol}) is a key to our forthcoming replacement of
the numerically awkward boundary conditions (\ref{bc}) by the much
more natural approximative truncation of recurrences
(\ref{recurrences}). We shall parallel the Hermitian construction
of ref. \cite{Classif} and try to bracket the exact energy between
its upper and lower estimates with $E \neq E(physical)$. Under
such an assumption our infinite series $\psi^{(ansatz)}(x)$ as
defined by equation (\ref{ansatz}) will {\em always} exhibit an
exponential asymptotic growth as described quantitatively by eq.
(\ref{asymp}) above. This means that we shall exempt the possible
lucky guess of the exact energy in its {\em full} precision as
{\em never} relevant in {\em any} step of our forthcoming
considerations. Such a very formal point of view does not
contradict the underlying physical intuition since boundary
conditions (\ref{bc}) are approximative. One has to move to the
limit $X_{R,L} \to \infty$ in principle.

The most important immediate consequence of our ``bracketing"
interpretation of boundary conditions is that at the large
absolute values of the coordinate $|x| \gg 1$ the first $N$
exponentially small components ${\cal O}(e^{-sx^2})$ may safely be
ignored as irrelevant. We may also insert (\ref{coeff}) and
(\ref{gensol}) in $ \psi^{(ansatz)}(x) \sim
\exp(-sx^2)\,\sum_{n=N+1}^{\infty}\, h_n \,(ix)^n$ with $N \gg 1$
and get
 \ben
 \psi^{(ansatz)}(x) \sim
e^{-sx^2}\,\sum_{n=N+1}^{\infty}\,
\frac{G_2\lambda^n(2)\,g_n(2)+G_5\lambda^n(5)\,g_n(5)
}{(3^{1/3})^n\,\Gamma(1+n/3)} \,(ix)^n, \ \ \ \ \ \ |x| \gg 1.
 \een
The validity of this formula is a strict consequence of the
specific constraint (\ref{constr}) imposed (say, from now on) upon
the admissible quasi-variational parameter $s$.

Once we split $ \psi^{(ansatz)}(x) =\psi^{(ansatz)}(G_2,G_5,x)$ in
its two components
 \ben
 \psi^{(ansatz)}(G_2,0,x) \sim G_2\,
e^{-sx^2}\,\sum_{n=N+1}^{\infty}\, \frac{(-x)^n\, \exp \left [
\gamma(2)\,n^{2/3} + {\cal O} (n^{1/3})
 \right ] }{(3^{1/3})^n\,\Gamma(1+n/3)},
 \label{bensa}
 \een
 \ben
 \psi^{(ansatz)}(0,G_5,x) \sim G_5\,
e^{-sx^2}\,\sum_{n=N+1}^{\infty}\, \frac{x^n\, \exp \left [
\gamma(5)\,n^{2/3} + {\cal O} (n^{1/3})
 \right ] }{(3^{1/3})^n\,\Gamma(1+n/3)},
 \label{cens}
 \een
we may apply the rule $e^z \sim (1+z/t)^t, \ t \gg 1$ in the error
term and get
 \ben
  \frac{ \psi^{(ansatz)}(G_2,0,-y)}
{\exp(-sy^2)}
  \sim G_2\,
\sum_{n=N+1}^{\infty} \frac{1}{(3^{1/3})^n\,\Gamma(1+n/3)}
 \left \{y \cdot
 \left [1+{\cal O}
 \left(\frac{1}{N^{1/3}}
 \right )
 \right ]
 \right \}^n
 \label{bsa}
 \een
and
 \ben
 \frac{
 \psi^{(ansatz)}(0,G_5,y)}
{\exp(-sy^2)}
  \sim G_5\,
\sum_{n=N+1}^{\infty} \frac{1}{(3^{1/3})^n\,\Gamma(1+n/3)}
 \left \{y \cdot
 \left [1+{\cal O}
 \left(\frac{1}{N^{1/3}}
 \right )
 \right ]
 \right \}^n.
 \label{cs}
 \een
This is valid at all the large arguments $y$. Along the {\em
positive} semi-axis $y \gg 1$, both the right-hand-side summands
are real and positive. They sum up to the same function
$\exp[y^3/3+ {\cal O}(y^2)]$. This is a consequence of the
approximation of the sum by an integral and its subsequent
evaluation by means of the saddle-point method. The same trick was
used by Hautot, in similar context, for the ${\cal P}-$symmetric
and Hermitian anharmonic oscillators \cite{Hautot}.

In contrast to the Hautot's resulting one-term estimates of
$\psi$, the present asymmetric, ${\cal PT}-$invariant construction
leads to the more general two-term asymptotic estimate
 \ben
 \psi^{(ansatz)}(G_2,G_5,x) \sim G_2\,
\exp[-x^3/3+ {\cal O}(x^2)]+
 G_5\, \exp[x^3/3+ {\cal O}(x^2)], \ \ \ \ |x| \gg 1.
 \label{finsa}
 \een
As long as we deal with the holomorphic function of $x$, this
estimate may be analytically continued off the real axis of $x$.
Near both the ends of the real line and within the asymptotic
wedges $|{\rm Im}\ x|/|{\rm Re}\ x| < \tan \pi/6$ we simply have
the rules
 \be
 \psi^{(ansatz)}(G_2,G_5,x) \sim G_2\,
\exp[-x^3/3+ {\cal O}(x^2)], \ \ \ \ {\rm Re}\ x <-X_L \ll - 1
 \label{leva}
 \ee
and
  \be
 \psi^{(ansatz)}(G_2,G_5,x) \sim G_5\,
\exp[x^3/3+ {\cal O}(x^2)], \ \ \ \ 1 \ll X_R <{\rm Re}\ x .
 \label{prava}
 \ee
They are fully compatible with formula (\ref{genere}) since $a=1$.

\section{The matrix form of the Hamiltonian}

Our complex differential Schr\"{o}dinger equation (\ref{SE})
becomes asymptotically real, in the leading-order approximation at
least. In a suitable normalization the wave functions
$\psi^{(ansatz)}(x)$ may be made asymptotically real as well. Near
infinity they will obey the standard Sturm Liouville oscillation
theorems \cite{Hille}. In particular, after a small decrease of
the tentative energy parameter $E> E(physical)$ the asymptotic
nodal zero $X_{R}$ or $-X_{L}$ originating in one of our boundary
conditions (\ref{bc}) will move towards infinity \cite{Fluegge}.

This may be re-phrased as follows. At a more or less correct
physical real pair $h_0=\rho\,\cos \zeta$ and $h_1=\rho\,\sin
\zeta$ with $\zeta \in (0, 2\pi)$ and with the convenient
normalization $\rho=1$ a small change of the energy $E$ somewhere
near its correct physical value $E_0 \approx E(physical)$ will
cause a sudden change of the sign of the asymptotically growing
exponentials (\ref{leva}) and (\ref{prava}) at some $\zeta_0
\approx \zeta(physical)$. This may be re-read as a doublet of
conditions
\be
G_2= G_2(E_0,\zeta_0)=0, \ \ \ \ \ \ G_5=G_5(E_0,\zeta_0) = 0.
\label{newbc}
 \ee
In the limit $N \to \infty$ of vanishing corrections, these two
requirements may be re-interpreted as a rigorous re-incarnation of
our original physical asymptotic boundary conditions (\ref{bc}).
The conclusion has several important consequences. Firstly, at a
fixed $N \gg 1$ we may define
 \ben f_p=G_p
\frac{\lambda^N(p)\,\exp [\gamma(p) N^{2/3}]}
{(3^{1/3})^N\,\Gamma(1+N/3)}, \ \ \ \ \ \ p = 2,5.
 \label{abbrevici}
 \een
Functions $f_p=f_p(E,\zeta_0)$ differ from their sign-changing
predecessors $G_p=G_p(E,\zeta_0)$ just by a constant factor near
$E_0$, $f_p(E,\zeta_0) \approx F_p \cdot (E-E_0)$. We may write
 \ben \ba h_N \approx
(F_2+F_5)(E-E_0)+{\cal O}[(E-E_0)^2],\\ (N+3)^{1/3}h_{N+1} \approx
[F_2\lambda(2)+F_5\lambda(5)](E-E_0) +{\cal O}[(E-E_0)^2] \ea
 \een
due to equation (\ref{coeff}).  This formula connects the two
functions $G_2,\,G_5$ with the two neighboring Taylor coefficients
$h_N=h_N(E_0,\zeta_0)$ and $h_{N+1}=h_{N+1}(E_0,\zeta_0)$ near the
physical $E_0$ and $\zeta_0$ by an easily invertible regular
mapping.  This means that the implicit algebraic boundary
conditions (\ref{newbc}) are strictly equivalent to the fully
explicit requirements
\be
h_N(E_0,\zeta_0) = 0, \ \ \ \ \ \ \ \ \ h_{N+1}(E_0,\zeta_0) = 0,
\ \ \ \ \ \ N \gg 1. \label{nec}
 \ee
By construction, this becomes an exact physical bound-state
condition in the limit $N \to \infty$. At the finite $N \gg 1$ its
appeal lies in its change-of-sign character. This need not make
equation (\ref{nec}) immediately suitable for computations but
once we fix $N=N_0\gg 1$, $E=E_0$, $\zeta=\zeta_0$ and insert the
zeros (\ref{nec}) in our recurrences (\ref{recurrences}), we
arrive at the truncated square-matrix equation
\be
  \left(
\begin{array}{ccccccc}
C_0&0&A_0&& & &\\ \delta&C_1&0&A_1&& & \\
\theta&\delta&\ddots&\ddots&\ddots&& \\ -\beta&\theta&\ddots&&&&
\\
 1&-\beta&\ddots&&&\ddots&A_{N-3}
\\
 &\ddots&\ddots&\ddots&\ddots&\ddots&0
\\
 &&1&-\beta&\theta&\delta&C_{N-1}
\\
\end{array} \right) \left( \ba
h_0\\ h_1\\ h_2\\ \ldots\\h_{N-3}\\ h_{N-2}\\ h_{N-1} \ea \right)
= 0. \label{4.8}
 \ee
This is our main result.  As long as $C_n=4sn+2s-E$, the energy
enters just the main diagonal and we may determine all its
approximate low-lying values $E_0$ by the routine $N \times
N-$dimensional diagonalization.

\section{Discussion}

\subsection{Illustrative numerical tests}

The smallest matrix in equation (\ref{4.8}) which contains all the
couplings has dimension $N=5$.  It is quite surprising that such a
drastic simplification leads to the mere 5\% or 6\% error in the
ground-state energy.  Together with the equally pleasant quick
increase of precision with the growing $N$, this is illustrated in
Table~1.  Table 2 shows where the numerical application of the
present approach can find its natural limitations. We observe a
steady decrease of precision at the higher excitations.

With $s=2$ and $a=c=\beta=\delta=1$, both Tables were computed in
MAPLE \cite{Maple}. This language keeps the possible loss of
precision under a careful control. This implies the growth of the
computing time at the higher dimensions. Still, the very quick
actual numerical rate of convergence enabled us to compute all our
examples on a current PC in a couple of minutes.

\subsection{Determinantal formulae for the Taylor coefficients}

With real $h_0$ and $h_1$ our wave functions $\psi^{(ansatz)}(x)$
are composed of the spacially symmetric real part and spacially
antisymmetric imaginary part.  Such a normalization fixes the
phases of the complex constants $G_2$ and $G_5$ accordingly, i.e.,
via equation (\ref{gensol}). This clarifies the structure of the
asymptotics of the wave functions.

Polynomial approximants of the Taylor series (\ref{ansatz}) offer
a reliable picture of $\psi(x)$ in a broad vicinity of the origin.
We may recall recurrences (\ref{recurrences}) and reveal that the
$h_0-$ and $h_1-$dependence of any coefficient $h_n$ is linear,
 \ben h_n=h_0\sigma_n+h_1\omega_n,
\ \ \ \ \ \ \ \ \ \ \sigma_0=\omega_1=1, \ \
 \sigma_1=\omega_0=0.
 \een
All three sequences $h_n$, $\sigma_n$ and $\omega_n$ satisfy the
same recurrences. As long as $\sigma_1=0$ and $ \omega_0=0$ we may
omit the second or first column from equation (\ref{4.8}) in the
latter two respective cases. In terms of the $(m+1)-$dimensional
matrices
 \ben \Sigma_m=
  \left(
\begin{array}{ccccccc}
C_0&A_0&& & &&\\ \delta&0&A_1&& & &\\ \vdots&C_2&0&A_2&&& \\
1&\vdots&C_3&\ddots&\ddots&&
\\
0&-\beta&\ddots&\ddots&0&A_{m-2}&
\\
\vdots &1&\ddots&\delta&C_{m-1}&0&A_{m-1}
\\
 &&\ddots&\ldots&\delta&C_m&0
\\
\end{array} \right)
\label{4.8cc}
 \een
and
 \ben \Omega_m=
  \left(
\begin{array}{ccccccc}
0&A_0&& & &&\\ C_1&0&A_1&& & &\\
\delta&C_2&0&A_2&&& \\ \vdots&\delta&C_3&\ddots&\ddots&&
\\
1&\vdots&\delta&\ddots&0&A_{m-2}&
\\
 &\ddots&-\beta&\ddots&C_{m-1}&0&A_{m-1}
\\
 &&1&\ldots&\delta&C_m&0
\\
\end{array} \right)
\label{4.8bb}
 \een
we may re-write not only the recurrences themselves but also their
unexpectedly compact solution
\be
\sigma_{n+1}=(-1)^{n}\frac{\det \Sigma_{n-1}}{n!(n+1)!}, \ \
\
\
\
\
 \omega_{n+1}=(-1)^{n}\frac{\det \Omega_{n-1}}{n!(n+1)!}, \ \ \
\ n = 1,2, \ldots .
 \ee
We need to know just the correct physical values of the three
variable parameters (viz., the norm $\rho=\sqrt{h_0^2+h_1^2}$, the
ratio $h_1/h_0 \equiv \tan \zeta$ and the physical energy $E$) in
order to be able to define our physical wave function $\psi(x)$
completely in terms of these closed formulae.

\subsection{Alternative ansatzs and constructions}

We have shown that the Taylor-series ansatz (\ref{ansatz})
mediates a useful transition from differential equation (\ref{SE})
to the difference equation (\ref{recurrences}), followed by its
further replacement by our final matrix Schr\"{o}dinger equation
(\ref{4.8}). In this context it is important to mention that our
choice of the initial form of ansatz (\ref{ansatz}) is by far not
unique.

A nice example of an alternative expansion may be found in ref.
\cite{ferdva} where the Hill-determinant study of the symmetric
potentials $V(x)=x^2 +\lambda\,u(x)$ with the non-polynomial
anharmonicity $u(x) = x^2/(1+g\,x^2)$ via the series of the form
(\ref{ansatz}) has been rendered possible by the use of the Taylor
series in powers of the ``adapted" variable $u(x)$. Sophisticated
versions of the latter trick move the (complex) singularities off
the physical domain of convergence and their active use in physics
dates back to Jaff\'{e} \cite{Jaffe} at least. They may even help
us to deal with relativistic corrections \cite{rela} etc.

Unfortunately, an application of the changes of variables to
asymmetric potentials is not without its specific difficulties.
Efficient methods of their suppression have been suggested,
therefore, in our older paper \cite{older} and, recently, by Bay
et al \cite{Karl}. Most often, one employs the {\em two
independent} separate ansatzs (one for each half-axis) and matches
the wave functions, say, in the origin.

In the latter comparison, the method of paper \cite{Karl} is most
straightforward. It is based simply on an introduction of the
second free parameter ($G$ or $V$ in the original notation). Even
from the very numerical point of view, the essence of the
algorithm of Bay et al remains purely iterative, therefore.

The more algebraic method of ref. \cite{older} works directly with
the matched, ``doubly infinite" sparse matrices. Although the
algorithm itself is already fully algebraized, its universality
seems redundant for our present purposes. Indeed, the ${\cal
PT}-$symmetric forces (\ref{AHO}) are only composed of the real
part which is spatially symmetric and of the non-vanishing
imaginary part which is spatially antisymmetric. This additional
information is well reflected and used by our present non-matching
approach.

We may summarize that we were able to preserve a maximal
similarity of our ``new Hill determinants" to their current
Hermitian predecessors (cf., e.g., ref. \cite{Estrin}). Moreover,
in a way completing the parallel studies of the other ${\cal PT}$
symmetric potentials, the very specific form of their spatial
asymmetry proved again ``extremely weak" from the purely
methodical point of view. We re-confirmed that its simplifying
role strongly resembles the role of the usual ${\cal P}-$symmetry
(i.e., parity), so useful in many parts of the current textbook
quantum mechanics.

\newpage

Table 1.

The $N-$dependence of energies.

$$
   \begin{array}{||c|ll||}
\hline
\hline
N&E_0&E_1\\
\hline
5& 1.793&7.547\\
6&1.823&5.868\\
7&1.634&5.856\\
8&1.673&5.138\\
9&1.627&5.162\\
10&1.658&4.922\\
15&1.693&5.106\\
20& 1.692&5.126\\
21& 1.691&5.124\\
22& 1.692&5.123\\
23& 1.692&5.123\\
24& 1.692&5.123\\
25& 1.692&5.123\\
\hline
\hline
\ea
$$

\newpage

\vspace{3cm}

Table 2. The
 growth of precision with dimension $N$.

$$
   \begin{array}{||cc|llllllll||}
\hline \hline &&\multicolumn{8}{|c||}{{ energies\ } E_n}\\
\hline
 &n&0&1&2&3&4&5&6&7\\ N&&&&&&&&&\\ \hline 15&& 1.69347&
    5.106&
    9.152&
    13.043&
    17.817&
    23.89&
    31.26&
    41.55\\
20&& 1.691638&
    5.12559&
    9.2800&
    14.050&
    19.244&
    -&
    32.35&
    -\\
25& &1.691579&
    5.123441&
    9.25812&
    13.8689&
    18.7925&
    24.265&
    30.039&
    37.97\\
30&& 1.691590&
    5.123614&
    9.26174&
    13.8826&
    18.8922&
    24.262&
    29.726&
    34.67\\
35& &1.691590&
    5.123579&
    9.26151&
    13.8793&
    18.8838&
    24.220&
    29.860&
    35.85\\
\hline
\hline
\ea
$$


\newpage
\begin{thebibliography}{99}

\bibitem{Weniger}
Bender C M and Wu T T 1969 Phys. Rev. 184 1231;

Killingbeck J 1980 J. Phys. A: Math. Gen. 13 49;

Simon B 1982 Int. J. Quant. Chem. 21 3;

Voros A 1983 Ann. Inst. Henri Poincar\'{e} A 39 211;

Turbiner A V 1989 Sov. Sci. Rev. A: Phys. 10 79;

Bender C M and Dunne G V 1989 Phys. Rev. D 40 3504;

Kleinert H 1990 Path Integrals in Quantum Mechanics, Statistics
and Polymer Physics (Singapore: World Scientific);

Arponen J S and Bishop R F 1990 Phys. Rev. Lett. 64 111;

Weniger E J 1996 Ann. Phys. (NY) 246 133;

Lay W 1997 J. Math. Phys. 38 639

\bibitem{BBjpa}
Bender C M and Boettcher S 1998 J. Phys. { A 31} L273

\bibitem{Ushveridze}
Ushveridze A G 1994 Quasi-Exactly Solvable Models in Quantum
Mechanics (Bristol: IOP)

\bibitem{Bessis}
Daniel Bessis 1992 private communication

\bibitem{Meis}
Le Guillou J C and Zinn-Justin J (ed) 1990 Large-Order Behaviour
of Perturbation Theory (Amsterdam: North Holland);

Bender C M and Milton K A 1998 Phys. Rev. D 57 3595

\bibitem{Navr}
Bishop R F, Flynn M F, Bosc\'{a} M C and Guardiola R 1989 Phys.
Rev. A 40 6154;

Navr\'{a}til P, Geyer H B and Kuo T T S 1993 Phys. Lett. B 315 1

\bibitem{Hatano}
Hatano N and Nelson D R 1996 Phys. Rev. Lett. 77 570;

Bender C M, Dunne G V and Meisinger P N 1998 preprint
cond-mat/9810369

\bibitem{Junker}
Cannata F, Junker G and Trost J 1998 Phys. Lett. { A 246}
 219;

Andrianov A A, Cannata F, Dedonder J P and Ioffe M V, preprint
quant-ph/9806019

\bibitem{Meisinger}
Bender C M, Boettcher S and Meisinger P N 1999 J. Math. Phys. 40
2201;

Fern\'andez F, Guardiola R,  Ros J and Znojil M 1999 J. Phys.{ A
32} 3105

\bibitem{BB}
Bender C M and Boettcher S 1998  Phys. Rev. Lett. { 24} 5243

\bibitem{mytri}
Fern\'andez F, Guardiola R,  Ros J and Znojil M 1998 J. Phys.{ A
31} 10105;

Bender C M and Dunne G V 1998 preprint quant-ph/9812039

\bibitem{Caliceti}
Calicetti E, Graffi S and Maioli M 1980 Commun. Math. Phys. 75 51;

Delabaere E and Pham F 1997 Ann. Phys. 261 180;

Blencowe M P, Jones H and Korte A P 1998 Phys. Rev. D 57 5092

\bibitem{Turbiner}
Bender C M and Turbiner A V 1993 Phys. Lett. { A 173} 442

\bibitem{Dyson}
Dyson F J 1952 Phys. Rev. 85 631

\bibitem{BBqed}
Bender C M and Milton K A 1999 J. Phys. { A 32} L87

\bibitem{Alvarez}
Alvarez G 1995 J. Phys. A: Math. Gen. 27 4589

\bibitem{list}
Richardson J L and Blankenbecler R 1979 Phys. Rev. D 19 496;

Halliday I G and Suranyi P 1980 Phys. Rev. D 21 1529;

Flessas G P and Watt A 1981 J. Phys. A: Math. Gen. 14 L315;

Chaudhuri R N 1985 Phys. Rev. D 31 2687;

Killingbeck J 1986 Phys. Lett. A 115 301;

Bishop R F and Flynn M F 1988 Phys. Rev. A 38 2211;

Hodgson R J W and Varshni Y P 1989  J. Phys. A: Math. Gen. 22 61;

Roychoudhury R K, Varshni Y P and Sengupta M 1990 Phys. Rev. A 42
184;

Bessis N and Bessis G 1997 J. Math. Phys. 38 5483;

Sergeev A V and Goodson D Z 1998 J. Phys. A: Math. Gen. 31 4301;

Sk\'{a}la L, \v{C}\'{\i}\v{z}ek J, Weniger E J and Zamastil J 1999
Phys. Rev. A 59 102

\bibitem{Ginsburg}
Ginsburg C A 1982 Phys. Rev. Lett. 48 839;

Tater M 1987 J. Phys. { A: Math. Gen. 20} 2483;

\bibitem{Estrin}
Estrin D A, Fernandez F M and Castro E A 1988 Phys. Lett. A 130
330;

Znojil M 1991 Phys. Lett. A 155 83

\bibitem{Singh}
Singh V, Biswas S N and Data K 1978 Phys. Rev. D 18 1901;

Cizek J and Vrscay E R 1984 Phys. Rev. A 30 1550;

Lee M H 1982 Phys. Rev. B 26 2547;

Masson D 1983 J. Math. Phys. 24 2974;

Lakhtakia A 1989 J. Phys. A: Math. Gen. 22 1791;

Arteca A G, Fernandez F M and Castro E A 1990 Large Order
Perturbation Theory and Summation Methods in Quantum Mechanics
(Berlin: Springer);

Meyer H-D, Hor\'{a}\v{c}ek J and Cederbaum L S 1991 Phys. Rev. A
43 3587

\bibitem{Graffi}
Graffi S and Grecchi V 1975 Lett. Nuovo Cimento 12 425;

Turchetti G 1978 Fortschr. Phys. 26 1;

Znojil M 1988 J. Math. Phys. 29 139;

Scherrer H, Risken H and Leiber T 1988 Phys. Rev. 38 3949;

Ahlbrandt C D 1996 J. Approx. Theory 84 188

\bibitem{Classif}
Znojil M 1994 J. Phys. { A: Math. Gen. 27} 4945

\bibitem{Birkhof}
Birkhoff G D 1930 Acta Math. 54 205

\bibitem{Karel}
Znojil M, Sandler K and Tater M 1985 J. Phys. { A: Math. Gen. 18}
2541

\bibitem{Karl}
Bay K, Lay W and Akopyan A 1997 J. Phys. { A: Math. Gen. 30} 3057
and further references quoted therein

\bibitem{soukr}
Znojil M 1988 J. Math. Phys. 29 1433

\bibitem{Hautot}
Hautot A 1986 Phys. Rev. D 33 437

\bibitem{Hille}
Hille E 1969 Lectures on Ordinary Differential Equations (Reading:
Addison-Wesley)

\bibitem{Fluegge}
Fl\"{u}gge S 1971 Practical Quantum Mechanics I (New York:
Springer) p. 153

\bibitem{Maple}
Char B W et al 1991 Maple V Language Reference Manual (New York:
Springer)

\v{C}\'{\i}\v{z}ek J, Vinette F and Weniger E J 1991 Int. J.
Quant. Chem. - Symp. 25 209 and  in de Groot R A and Nadrchal J
(ed) 1993 Physics Computing '92 (Singapore: World Scientific), p.
31

\bibitem{ferdva}
Fernandez F M 1991 Phys. Lett. { A 160} 116

\bibitem{Jaffe}
Jaff\'{e} G 1933 Z. Phys. { 87} 535;

Solov'ev E A 1981 Sov. Phys. - JETP 54 893

\bibitem{rela}
Znojil M 1996 J. Phys. { A: Math. Gen. 29} 2905

\bibitem{older}
Znojil M 1992 J. Math. Phys. 33 213


\end{thebibliography}
\end{document}